\documentclass[seceq]{ptptex}

\usepackage{graphicx}

\def\simge{\mathrel{%
       \rlap{\raise 0.511ex \hbox{$>$}}{\lower 0.511ex \hbox{$\sim$}}}}
\def\simle{\mathrel{
       \rlap{\raise 0.511ex \hbox{$<$}}{\lower 0.511ex \hbox{$\sim$}}}}

\title{
Lattice QCD at finite temperature and density%
}

\author{
Shinji \textsc{Ejiri}%
}

\inst{
Graduate School of Science and Technology, Niigata University, \\
Niigata 950-2181, Japan
}

\abst{
We study the phase structure of QCD at finite temperature and density 
by numerical simulations on a lattice. 
The most important point for the numerical study at finite density is 
treatment of the sign problem. 
We propose a method to avoid the sign problem, which is based on a cumulant expansion of the complex phase in the density of state method combined with the reweighting method. 
Using the method, we study the critical point terminating a first order 
phase transition line in lattice QCD at high temperature and density.
}

\begin{document}

\maketitle

\section{Introduction}
\label{sec:intro}

It is very important to explore the QCD phase structure to understand the history of the universe.
We expect that the nature of the chiral phase transition changes as the quark number density increases. Moreover, new state of QCD matter may appear at high density. 
However, because the quark determinant is complex at finite density, the Monte-Carlo method is not applicable directly for finite density QCD.
One of the popular methods to avoid this problem is the density of state method.
We adopt an appropriate physical quantity such as quark number, chiral order parameter, gauge action etc., which is denoted by $X$, and discuss its state density. The state density, i.e. the probability distribution function, at finite temperature $T$ and quark chemical potential $\mu_q$ is defined by
\begin{eqnarray}
W(X',T,\mu_q) 
= \int {\cal D} U \ \delta(X'-X) \ (\det M)^{N_{\rm f}} e^{-S_g}, 
\label{eq:xdist}
\end{eqnarray}
where $\delta(x)$ is the delta function, $M$ is the quark matrix,  
$S_g$ is the gauge action, and $N_{\rm f}$ is the number of flavors. 
The partition function ${\cal Z}$ is given by 
${\cal Z} (T, \mu_q)= \int W(X,T,\mu_q) dX$.
Once we obtain the probability (\ref{eq:xdist}), expectation values of the operator ${\cal O}$ of $X$, 
e.g. $\langle X \rangle, \langle (X^2 - \langle X \rangle )^2 \rangle$, 
can be evaluated by the following equation;
\begin{eqnarray}
\langle {\cal O}[X] \rangle = \frac{1}{\cal Z} \int {\cal O}[X] W(X) dX. 
\label{eq:xexpe}
\end{eqnarray}
Although these equations are rather trivial, the probability distribution of a physical quantity, Eq.~(\ref{eq:xdist}), is well-defined as a real number even when $\det M$ is complex. 

In this report, we discuss the density of state approach combined with the reweighting method to investigate the QCD phase structure at high density.
In the next section, we discuss the state density using the reweighting method.
For this calculation, we introduce a method to avoid the sign problem in Sec.~\ref{sec:phase}. 
Using the method, we study the phase diagram. 
The investigation of the distribution function is one of the most primitive approaches to identify the order of phase transitions.
We expect that two phases coexist at a first order phase transition point. 
In Sec.~\ref{sec:appl}, we calculate the distribution function and discuss the order of phase transitions using the distribution function. 
Conclusions are given in Sec.~\ref{sec:conc}.

\section{Density of state in the reweighting method}
\label{sec:dosmethod}

\begin{figure}[tb]
  \begin{center}
    \begin{tabular}{cc}
      \includegraphics[width=65mm]{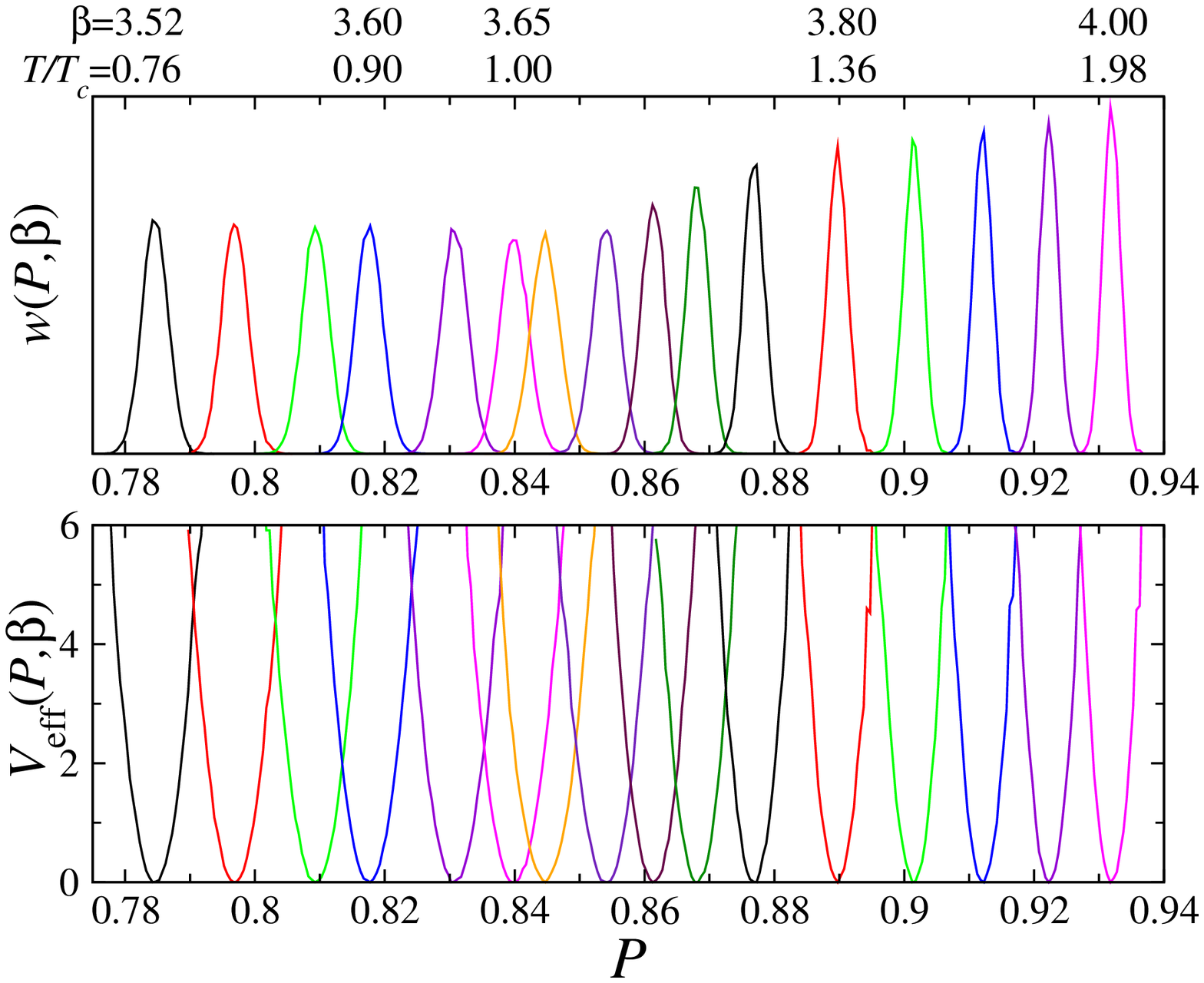} &
      \includegraphics[width=60mm]{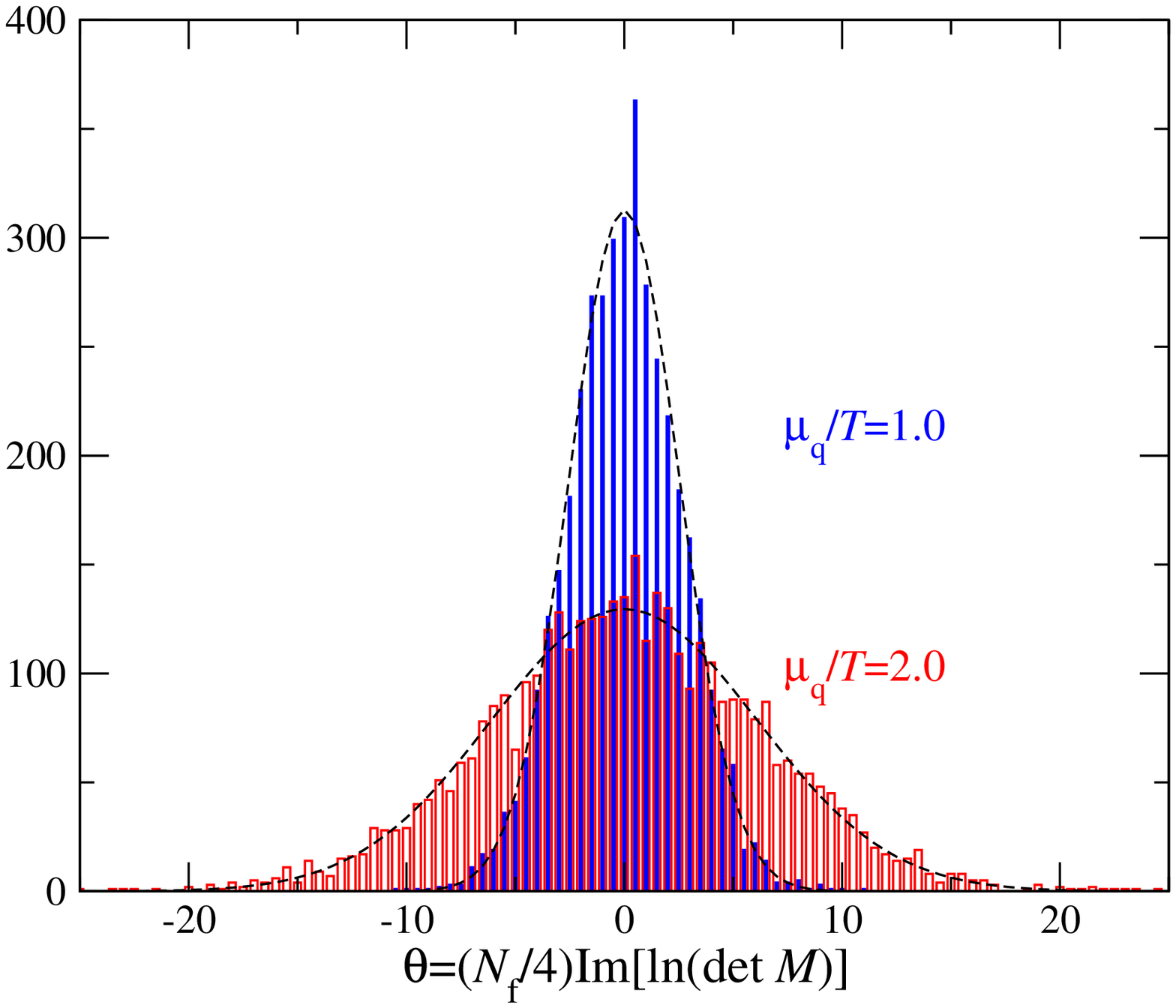}
    \end{tabular}
    \vspace*{-2mm}
    \caption{Left: The plaquette histogram $w(P,\beta)$ and the effective potential $V_{\rm eff}(P,\beta,0)$ at $\mu_q=0$ for each $\beta$.
    Right: The histograms of complex phase $\theta$ for $\mu_q/T= 1.0$ and $2.0$ at $\beta=3.65$\cite{eji07}.
    }
    \label{fig1}
  \end{center}
\end{figure}

We discuss the density of state method with fixing the plaquette variable $(P)$,
i.e. $1 \times 1$ Wilson loop, as an example. The distribution function is defined by Eq.~(\ref{eq:xdist}) with $X=P$.
For later discussions, we define the average plaquette $P$ as 
$P \equiv -S_g/(6 \beta N_{\rm site})$ 
and the quark matrix $M$ as independent of $\beta$. 
$N_{\rm site}$ is the number of sites, and 
the parameter $\beta \equiv 6/g^2$ controls the temperature.
When the quark determinant is real, the distribution function is 
given by the histogram of $P$. 
The plaquette distribution function and $V_{\rm eff}(P) \equiv - \ln W(P)$ for p4-improved staggered fermions at $\mu_q=0$ obtained in Ref.~\citen{BS05} are shown in Fig.~\ref{fig1} (left). $V_{\rm eff}(P)$ is normalized at the minimum point, and the minimum point moves to right as the temperature $(T/T_c)$ or $\beta$ increases. 
$T_c$ is the transition temperature at $\mu_q=0$.
We denote the distribution function at $\mu_q=0$ as $ w(P,\beta) \equiv W(P,\beta,0)$.

Because the quark determinant is complex at finite $\mu_q$, the reweighting method is used to obtain the distribution function\cite{eji07}. 
The partition function is rewritten as 
\begin{eqnarray}
{\cal Z}(\beta, \mu_q) 
= \int W(P,\beta,\mu_q) \ dP
= \int R(P,\mu_q) w(P,\beta) \ dP.
\label{eq:rewmu}
\end{eqnarray}
Here, $R(P,\mu_q)$ is the reweighting factor for finite $\mu_q$ 
defined by 
\begin{eqnarray}
R(P',\mu_q) \equiv 
\frac{\int {\cal D} U \delta(P'-P) (\det M(\mu_q))^{N_{\rm f}}}{
\int {\cal D} U \delta(P'-P) (\det M(0))^{N_{\rm f}}} 
= \frac{ \left\langle \delta(P'-P) 
\frac{(\det M(\mu_q))^{N_{\rm f}}}{(\det M(0))^{N_{\rm f}}} 
\right\rangle_{(\beta, \mu_q=0)} }{
\left\langle \delta(P'-P) \right\rangle_{(\beta, \mu_q=0)}}. \hspace{8mm}
\label{eq:rmudef}
\end{eqnarray}
This $R(P, \mu_q)$ is independent of $\beta$, and 
$R(P, \mu_q)$ can be measured at any $\beta$. In this method, 
all simulations are performed at $\mu_q=0$ and the effect of 
finite $\mu_q$ is introduced through the operator 
$\det M(\mu_q) / \det M(0)$ measured on the configurations 
generated by the simulations at $\mu_q=0$.

Since QCD has the symmetry of charge conjugation, 
the partition function is invariant under a change from 
$\mu_q$ to $-\mu_q$, i.e. $R(P,-\mu_q) = R(P,\mu_q)$.
Moreover, the quark determinant satisfies $\det M (-\mu_q) = (\det M(\mu_q^*))^*$.
From these equations, we get 
$[R(P, \mu_q)]^* = R(P, \mu_q^*)$.
This indicates that $R(P, \mu_q)$ is real if $\mu_q$ is real, 
i.e. $\mu_q = \mu_q^*$, and the probability distribution function of the plaquette given by 
$W(P, \beta, \mu_q) = R(P, \mu_q) w(P, \beta)$ is real.

\section{Avoiding the sign problem}
\label{sec:phase}

However, a serious sign problem occurs in the calculation of $R$ for large $\mu_q/T$. 
The histogram of the complex phase $\theta$ are shown in Fig.~\ref{fig1} (right) obtained in a simulation by p4-improved staggered fermions \cite{eji07}. 
The complex phase of the quark determinant is defined by a Taylor expansion;
\begin{eqnarray}
\theta (\mu) 
& = & N_{\rm f} {\rm Im} \,[\ln \det M(\mu_q)]
\nonumber\\
&=& N_{\rm f} \sum_{n=0}^{\infty} \frac{1}{(2n+1)!} {\rm Im}
\left[ \frac{\partial^{2n+1} (\ln \det M(\mu_q))}{\partial (\mu_q/T)^{2n+1}} 
\right]_{(\mu_q=0)} (\mu_q/T)^{2n+1} .
\label{eq:tatheta}
\end{eqnarray}
We note that $\ln \det M(\mu_q)$ is not uniquely defined for complex $\det M(\mu_q)$.
The $\theta$ defined in Eq.~(\ref{eq:tatheta}) is not restricted to be in the range $-\pi$ to $\pi$, and the maximum value of $|\theta|$ is infinite in the large volume limit. 
Of course, we can restrict the range of $\theta$ from $-\pi$ to $\pi$ subtracting $2 \pi n$, where $n$ is an integer, in the definition of $\theta$. However, this ambiguity does not affect the calculation of $\langle e^{i \theta} \rangle$.

In Fig.~\ref{fig1} (right), the width of the distribution becomes wider as $\mu_q/T$ increases, corresponding to the phase fluctuation larger.
The expectation value of $\langle e^{i \theta} |\det M| \rangle$ decreases as the fluctuation of $\theta$ increases, and the expectation value becomes smaller than the statistical error when the complex phase fluctuation of the quark determinant becomes larger than $O(\pi)$ in the Monte-Carlo steps. 
This is the sign problem in the calculation of the reweighting factor. 

To avoid the sign problem, we perform the $\theta$ integration before the integration of $| \det M(\mu_q)/ \det M(0) |^{N_{\rm f}} \equiv F$ 
in the calculation of Eq.~(\ref{eq:rmudef}); 
$\langle e^{i \theta} F \rangle = \int \langle e^{i \theta} \rangle_F \ F \ dF$,
where $\langle \cdots \rangle_F$ means the expectation value with fixed $F$.
We then consider the following cumulant expansion; 
\begin{eqnarray}
\langle e^{i \theta} \rangle_F = 
\exp \left[i \left\langle \theta \right\rangle_c
- \frac{\langle \theta^2 \rangle_c}{2} 
- \frac{i \left\langle \theta^3 \right\rangle_c}{3!} 
+ \frac{\langle \theta^4 \rangle_c }{4!} 
+ \frac{i \langle \theta^5 \rangle_c }{5!} 
- \frac{\langle \theta^6 \rangle_c}{6!} + \cdots \right],
\label{eq:cum}
\end{eqnarray}
where $\langle \theta^n \rangle_c$ is the $n^{\rm th}$ order cumulant, e.g.
$
\left\langle \theta^2 \right\rangle_c 
= \left\langle \theta^2 \right\rangle_F , 
\hspace{3mm}
\left\langle \theta^4 \right\rangle_c 
= \left\langle \theta^4 \right\rangle_F
-3 \left\langle \theta^2 \right\rangle_F^2, 
\hspace{3mm}
\left\langle \theta^6 \right\rangle_c 
= \left\langle \theta^6 \right\rangle_F
-15 \left\langle \theta^4 \right\rangle_F 
\left\langle \theta^2 \right\rangle_F 
+30 \left\langle \theta^2 \right\rangle_F^3 .
\hspace{3mm}
$
Note that $\langle \theta^n \rangle_c =0$ for odd $n$
due to the symmetry under $\theta \rightarrow -\theta$. 
Because only the odd-order cumulants are the source of the complex phase in 
$\langle \exp(i \theta) \rangle_F$, the value of 
$\langle \exp(i \theta) \rangle_F$ is guaranteed to be real and positive 
from this symmetry if the cumulant expansion converges. 
Although the identity (\ref{eq:cum}) is exact if we consider infinite terms of the expansion, there is no source of the sign problem once we eliminate the odd terms.

As shown in Fig.~\ref{fig1} (right), the distribution of $\theta$ is well-approximated by a Gaussian function (dashed line). 
When the distribution of $\theta$ is Gaussian, the $O(\theta^n)$ terms vanish for $n >2$ in Eq.~(\ref{eq:cum}). 
Hence, the approximation that the higher order cumulants are neglected except for the first nonzero term is equivalent to the Gaussian approximation for the $\theta$ distribution.
When one wants to improve the Gaussian approximation, it is achieved 
by adding higher order terms.

Moreover, the cumulant expansion can be regarded as a power expansion 
in terms of $\mu_q$ because $\theta \sim O(\mu_q)$. 
Therefore, if we take into account the cumulants up to the $n^{\rm th}$ order, 
the truncation error does not affect the Taylor expansion up to $O(\mu_q^n)$.
The Gaussian approximation corresponds to the leading non-trivial 
order approximation of the Taylor expansion in $\mu_q$.

On the other hand, a careful discussion about the infinite volume 
$(V)$ limit is required\cite{whot10}.
Because the operator $\theta$ is roughly proportional to $V$, 
the $n^{\rm th}$ order cumulant $\langle \theta^n \rangle_c$ may 
increase as $O(V^n)$ naively.
If this is the case, the cumulant expansion does not converge at large $V$.
However, the following argument suggests that the convergence property of the cumulant expansion is independent of the volume when the correlation length of the system is finite. 
Note that, since no critical point is expected to exist in two-flavor QCD 
at $m_q > 0$ and $\mu_q=0$, the correlation length between quarks is finite.
The expansion coefficients of $\theta$ in Eq.~(\ref{eq:tatheta}) are 
given by combinations of traces of products of $M^{-1}$, 
$\partial^n M/ \partial (\mu_q/T)^n$ and so on.
For example, the first coefficient is given by the trace of 
$N_f [M^{-1} (\partial M/ \partial (\mu_q/T))]$ 
and the diagonal element of this matrix is the local quark number density
operator $(\sim \bar{\psi} \gamma_0 \psi(x))$ at $\mu_q=0$. 
When the correlation length of the local number density operator is much shorter than the system size, we may decompose the first derivative term into independent contributions from spatially separated regions. 
The same discussion is applicable to higher order coefficients too. 

In this case, one can write the phase as $\theta=\sum_x \theta_x$, 
where $\theta_x$ is the contribution from a spatial region labeled by $x$ 
and these contributions are independent.
The average of $\exp(i\theta)$ is thus 
\begin{eqnarray}
\left\langle e^{i\theta} \right\rangle 
\approx \prod_x \left\langle e^{i\theta_x} \right\rangle 
= \exp \left( \sum_x \sum_n \frac{i^n}{n!} 
\left\langle \theta_x^n \right\rangle_c \right).
\end{eqnarray}
This equation suggests that all cumulants $\langle \theta^n \rangle_c 
\approx \sum_x \left\langle \theta_x^n \right\rangle_c$ 
increase in proportion to the volume as the volume increases. 
Therefore, while the width of the distribution, i.e. the phase fluctuation, 
increases in proportion to the volume, 
the ratios of the cumulants are independent of the volume.
The higher order terms in the cumulant expansion are well under control in the large volume limit.

In addition, the complex phase can be decomposed into independent parts 
when we define $\theta$ as 
$\theta= \int_0^{\mu_q/T} 
{\rm Im} [d \ln \det M / d(\mu_q/T)]_{(\mu_q'/T)} d(\mu_q'/T)$,
as well as Eq.~(\ref{eq:tatheta}).

Because $\theta$ is $O(\mu_q)$ and $\langle \theta^n \rangle_c$ is $O(\mu_q^n)$, 
the Gaussian approximation is valid at small $\mu_q$ and 
the higher order cumulants will become visible at large $\mu_q$. 
The application range of the Gaussian approximation in terms of $\mu_q$ must 
be checked for each analysis by calculating the ratio of cumulants. 
However, it is expected from the argument of the volume-dependence of the ratios 
that the application range does not change once the system size becomes 
larger than the correlation length.
This property will enable us to use large lattices.

\section{Distribution function and first order phase transition}
\label{sec:appl}

\begin{figure}[tb]
  \begin{center}
    \begin{tabular}{cc}
      \includegraphics[width=65mm]{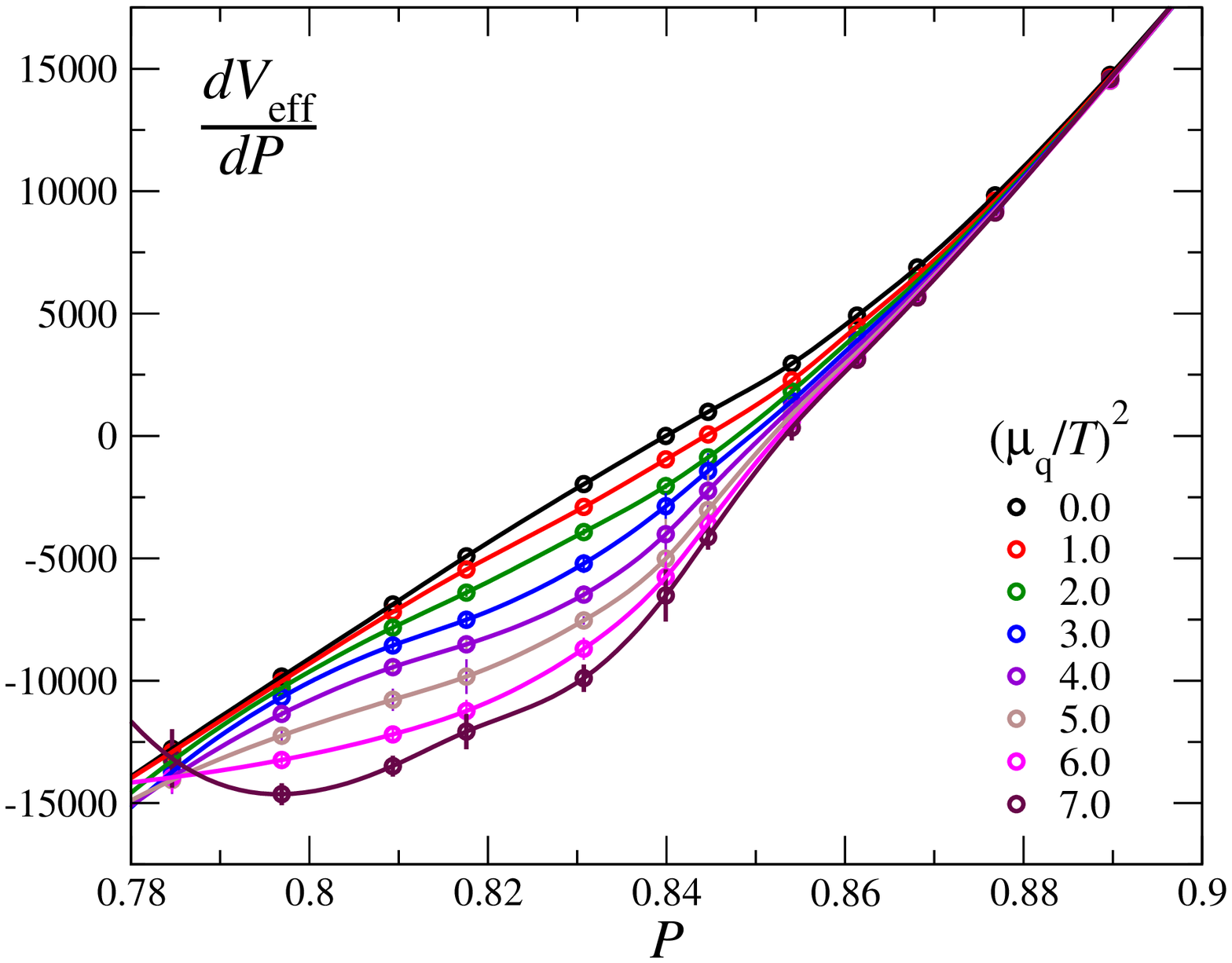} &
      \includegraphics[width=65mm]{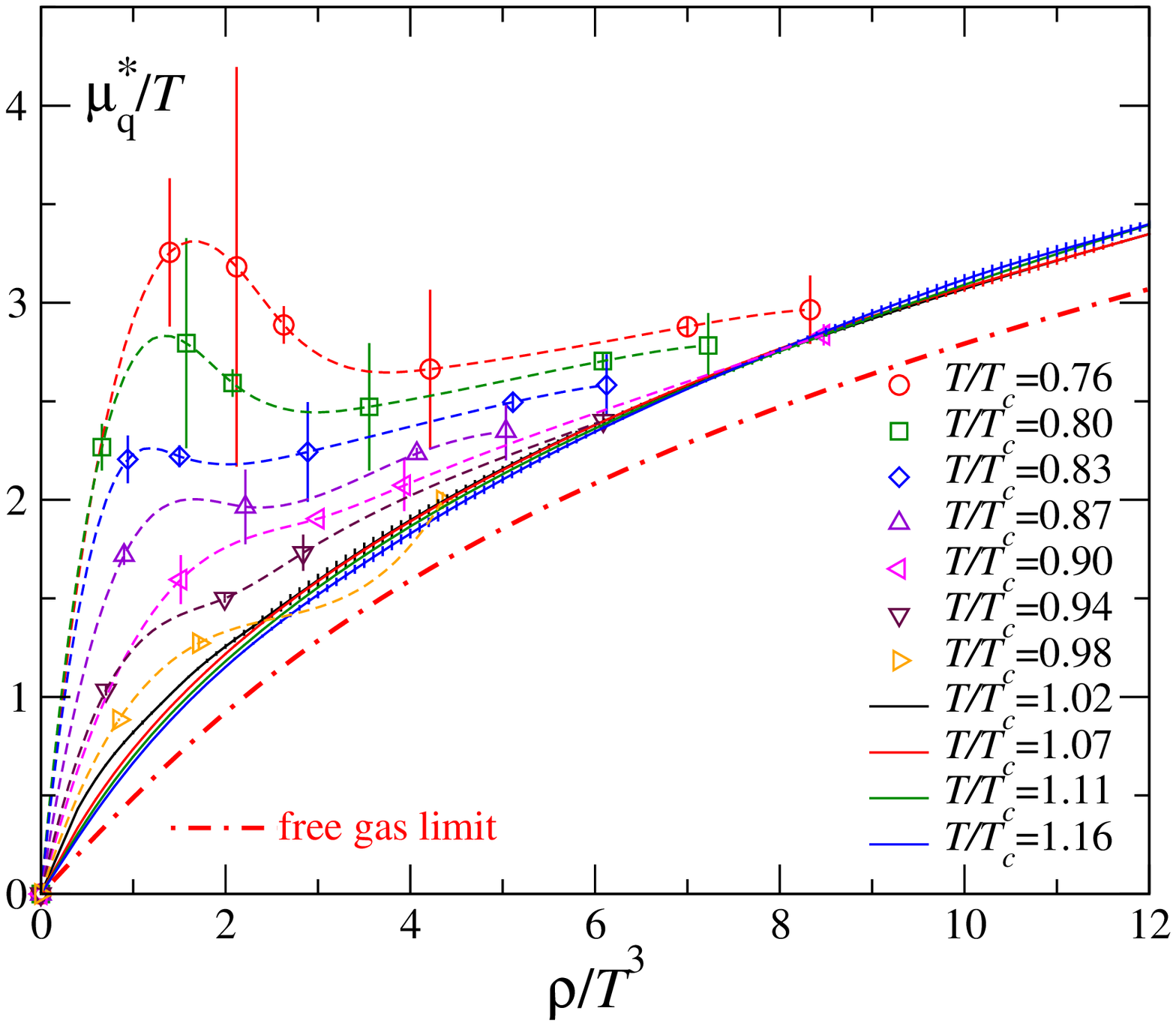}
    \end{tabular}
    \vspace*{-3mm}
    \caption{Left: 
The derivative of the effective potential $dV_{\rm eff}/dP$ at $\beta=3.65$\cite{lyz09}. 
We measured them at the peaks of the plaquette histograms in Fig.~{\protect \ref{fig1}} (left) and interpolated the data by a cubic spline method.
    Right: The derivative of $\ln {\cal Z}_C$ vs. quark number density computed with a saddle point approximation \cite{eji08}. $T_c$ is the transition temperature at $\mu_q=0$.
    }
    \label{fig2}
  \end{center}
\end{figure}

In this section, we discuss the order of the phase transition at finite density \cite{eji07,lyz09} using data obtained in simulations with the 2-flavor p4-improved staggered quarks \cite{BS05}. $m_{\pi} \approx 770 {\rm MeV}$.
The distribution function is expected to be a double-peaked function at a first order transition point, i.e. $V_{\rm eff} = - \ln W$ is a double-well function. 
It is easy to prove useful properties in the plaquette effective potential;
From the definition of $W$ and $P$, 
$ W(P, \beta, \mu_q) 
= e^{6 (\beta - \beta_0) N_{\rm site} P} W(P, \beta_0, \mu_q)$
is satisfied under the parameter change from $\beta_0$ to $\beta$. Then, 
$d V_{\rm eff}/dP$ at different $\beta$ can be estimated by the equation; 
\begin{eqnarray}
\frac{d V_{\rm eff}}{dP} (P,\beta, \mu_q)
= \frac{d V_{\rm eff}}{dP} (P,\beta_0, \mu_q) -6 (\beta - \beta_0) N_{\rm site},
\label{eq:derrewbeta}
\end{eqnarray}
and $d^2 V_{\rm eff}/dP^2$ is independent of $\beta$.
Therefore, the shape of $dV_{\rm eff}/dP$ as a function of $P$ does not change 
with $\beta$ up to a $\beta$-dependent constant.

Using Eq.~(\ref{eq:derrewbeta}), we calculate $dV_{\rm eff}/dP$ at $\mu_q=0$ in a wide range of $P$.
Performing simulations at many $\beta$ and finding the peak position of the plaquette distribution, at which $dV_{\rm eff}/dP (P) =0$, the value of $dV_{\rm eff}/dP (P)$ for $\beta_0$ is given by $6 (\beta - \beta_0) N_{\rm site}$. 
This method is much easier than the estimation from the plaquette histogram because the range of plaquette value obtained by a simulation with single $\beta$ is narrow. 
(See Fig.~\ref{fig1} (left).)
The finite density effect of $V_{\rm eff}$ is evaluated calculating $R(P, \mu_q)$
with the Gaussian approximation.
We plot $d V_{\rm eff}/dP$ instead of $V_{\rm eff}$ itself in Fig.~\ref{fig2} (left) for various $\mu_q/T$.
$\beta=3.65$ is adopted for these results, however the $\beta$ value can be easily changed by Eq.~(\ref{eq:derrewbeta}). 
If the effective potential $V_{\rm eff}$ is a double-well function of $P$, there exists a region of $P$ where the derivative of $ d V_{\rm eff}/d P$ is negative. 
The left panel of Fig.~\ref{fig2} shows that the region of $d^2 V_{\rm eff}/d P^2 <0$ exists for $(\mu_q/T)^2 > 6$. This suggests that the phase transition becomes first order at high density. The details of this analysis are given in Refs.~\citen{eji07,lyz09}.

Finally, we want to mention the distribution function of the quark number \cite{eji08}.
The probability distribution is, in principle, measurable by event-by-event analysis of heavy-ion collisions. 
The Gaussian approximation is also useful for the calculation of the quark number distribution function.
The relation between the grand canonical partition function ${\cal Z}$ and the canonical partition function ${\cal Z}_{\rm C}$ is given by the following Laplace transformation;
\begin{eqnarray}
{\cal Z}(T,\mu_q) 
= \sum_{N} \ {\cal Z}_{\rm C}(T,N) e^{N \mu_q/T}
= V \int \ {\cal Z}_{\rm C}(T, \rho V) e^{\rho V \mu_q/T} d \rho. 
\label{eq:cpartition} 
\end{eqnarray}
where $N$ is the quark number, $V$ is the volume and $\rho \equiv N/V$ is the quark number density.
$- \ln {\cal Z}_{\rm C} - \rho V \mu_q/T$ is regarded as the effective potential $V_{\rm eff}(\rho)$.

We compute the derivative of $\ln {\cal Z}_C$ with respect to $\rho$ by the saddle point approximation using the data obtained in Ref.~\citen{BS05}. 
If the distribution function is a double-peaked function, the derivative of $\ln {\cal Z}_C$ is an S-shaped function. 
Here, we denote $\mu^*/T \equiv -(1/V) d \ln {\cal Z}_C / d \rho$, since $\mu^*/T = \mu/T$ in the thermodynamic limit.
This calculation suffers from the sign problem. 
To eliminate the sign problem, the approximation discussed in the previous section is used, i.e. the complex phase factor $e^{i \theta}$ is replaced by $\exp[- \langle \theta^2 \rangle /2]$. 
The details are given in Ref.~\citen{eji08}.

The result of $\mu_q^*/T$ is shown in Fig.~\ref{fig2} (right) as a function of $\rho/T^3$ for each temperature $T/T_c$. 
The dot-dashed line is the value of the free quark-gluon gas in 
the continuum theory, 
$\rho/T^3 = N_{\rm f} [ (\mu_q/T) + (1/\pi^2) (\mu_q/T)^3]$.
From this figure, we find that a qualitative feature of $\mu_q^*/T$ 
changes around $T/T_c \sim 0.8$, i.e. $\mu_q^*/T$ increases monotonically 
as $\rho$ increases above 0.8, whereas it shows an S-shape below 0.8. 
The behavior at low $T$ is a signature of a first order phase transition. 
Although some approximations are used, the critical value of $T$ is roughly consistent with the critical point estimated by the plaquette effective potential using the same configurations, $(T/T_c, \mu_q/T) \approx (0.76, 2.5)$ \cite{eji07}. 
The difference between these two results may be a systematic error. 
Further studies are necessary to predict the critical point quantitatively, but these results are consistent with our qualitative expectation.

\section{Summary}
\label{sec:conc} 

We discussed methods to investigate finite density QCD beyond the low density region. 
A method based on the investigation of an effective potential as a function of the average plaquette was proposed introducing an approximation to avoid the sign problem, and the existence of the critical point at finite density is suggested by simulations with improved staggered quarks.
Moreover, it was found that interesting information about the QCD phase structure at finite density is obtained by constructing the canonical partition function for each quark number.


%


\begin{thebibliography}{99}

\bibitem{BS05}
C.R. Allton, M. D\"{o}ring, S. Ejiri, S.J. Hands, O. Kaczmarek, F. Karsch, 
E. Laermann and K. Redlich, \PRD{71,2005,054508}.

\bibitem{eji07}
S. Ejiri, \PRD{77,2008,014508}. 

\bibitem{whot10} 
S. Ejiri, Y. Maezawa, N. Ukita, S. Aoki, T. Hatsuda, N. Ishii, K. Kanaya 
and T. Umeda (WHOT-QCD Collaboration), \PRD{82,2010,014508}.

\bibitem{lyz09} 
S. Ejiri and H. Yoneyama, PoS \textbf{LAT2009} (2009), 173.

\bibitem{eji08}
S. Ejiri, \PRD{78,2008,074507}.

\end{thebibliography}
\end{document}